{cirant1.eps}
{cirant1.eps}
\documentclass[letterpaper,11 pt]{article}

\usepackage{epsfig,xspace,amsmath,amssymb,pifont,wrapfig,bbm}
\newcommand{\cC}{{\mathcal C}} \newcommand{\Cc}{$\cC$\xspace}
\newcommand{\cD}{{\mathcal D}} \newcommand{\Dc}{$\cD$\xspace}
\newcommand{\cS}{{\mathcal S}} \newcommand{\Sc}{$\cS$\xspace}
\newcommand{\cR}{{\mathcal R}} 
 
\newcommand{\cQ}{{\mathcal Q}} \newcommand{\Qc}{$\cQ$\xspace}
\newcommand{\cU}{{\mathcal U}} 

\newcommand{\cUop}{\cU^\ast}\newcommand{\Ucop}{$\cUop$\xspace}
\newcommand{\cQop}{\cQ^\ast}\newcommand{\Qcop}{$\cQop$\xspace}

\newcommand{\compat}{\,\heartsuit}
\newcommand{\Compat}{\heartsuit}
\newcommand{\fotw}{\frac{1}{2}}
\newcommand{\fotr}{\frac{1}{3}}

\newtheorem{theorem}{Theorem}
\newtheorem{lemma}{Lemma}
\newtheorem{thm}{Theorem}[section]

\newtheorem{cor}[thm]{Corollary}

\newenvironment{proof}{\noindent{\bf Proof}. }{\hfill\ding{113}\\[0.5ex]}

\newcommand{\cI}{{\mathcal I}} 
\newcommand{\eps}{\varepsilon}
\newcommand{\es}{\varnothing}

\newcommand{\IE}{{\em i.e.}\xspace}
\newcommand{\MAV}{{\sc MinAntVar}\xspace}

\newcommand{\MAa}{{\sc MinAnt}\xspace}

\newcommand{\MAb}{{\sc MinAntLoad}\xspace}
\newcommand{\Parti}{{\sc Partition}\xspace}

\newcommand{\Bab}{{\sc BinSchedule}\xspace}
\begin{document}
\title{Exact and Approximation Algorithms for 
Geometric and Capacitated Set Cover Croblems
with Applications
\thanks{Research supported in part by 
DFG grants, the Hausdorff Center research grant EXC59-1, and
the VR grant 621-2005-4085.}}
\author{Piotr Berman 
\thanks{Department of Computer Science and Engineering, 
Pennsylvania State University.
Email: berman@cse.psu.edu.}
\and 
Marek Karpinski 
\thanks{Department of Computer Science, Bonn University.
Email: marek@cs.uni-bonn.de}
\and
Andrzej Lingas
\thanks{Department of Computer Science,
Lund University.
Email: Andrzej.Lingas@cs.lth.se}}
\date{}
\maketitle
\begin{abstract}
{\small
First,
we study geometric variants of the standard set
cover motivated by assignment of directional 
antenna and shipping with deadlines,
providing the first known polynomial-time exact solutions.

Next, we consider
the following general
capacitated set cover problem.
There is given a set of elements with real weights
and a family {\cal S} of sets of elements.  One can use a set
if it is a subset of one of the sets on our lists and the sum of weights is at
most one.  The goal is to cover all the elements with the 
allowed sets.

We show that any polynomial-time algorithm that approximates 
the un-capacitated version of
the set cover problem with ratio $r$ can be converted to an approximation
algorithm for the capacitated version with ratio $r+1.357$.

In particular,
the composition of these two results yields a polynomial-time approximation
algorithm for the problem of covering a set of customers
represented by a weighted $n$-point set with a minimum
number of antennas of variable angular range and
fixed capacity with ratio $2.357.$

Finally, we provide a PTAS for the dual problem where
the number of sets (e.g., antennas) to use is fixed and the
task is to minimize the maximum set load, in case
the sets correspond to line intervals or arcs.

}
\end{abstract}


\newpage

\section{Introduction}
In this paper, we study special geometric set cover
problems and capacitated set cover problems.

In particular, the shapes of geometric sets we
consider correspond to those of potential
directional antenna ranges. Several geometric
covering problems where  a planar point set
is to be covered with a minimum number
of objects of a given shape have been studied
in the literature, e.g., in \cite{goodrich,clarkson,hoch1}.

On the other hand, a capacitated set cover problem can be seen as a
generalization of the classical bin packing problem (e.g., see
\cite{Cof}) to include several types of bins. Thus, we are given a set
of elements $\{1,\ldots,n\}$, each with a demand $d_i,$ and a set of
subsets of $\{1,\ldots,n\}$ (equivalently, types of bins), 
and the goal is to partition the elements
into a minimum number of copies of the subsets (bins) so the total demand of
elements assigned to each set copy does not exceed a fixed upper
bound $d.$

Capacitated set cover problems are useful abstraction in
studying the problems of minimizing the number of directional
antennas. The use of directional antennas in cellular and wireless
communication networks steadily grows
\cite{bao02,CPeraki03,SYi03,spyropoulos02}.  Although such
antennas can only transmit along a narrow beam in a particular
direction they have a number of advantages over the standard
ones. Thus, they allow for an additional independent communication
between the nodes in parallel \cite{spyropoulos02}, they also attain
higher throughput, lower interference, and better energy-efficiency
\cite{bao02,CPeraki03,SYi03}.

\begin{figure}[ht]
\begin{center}
\vspace{2ex}
\includegraphics[scale=0.75]{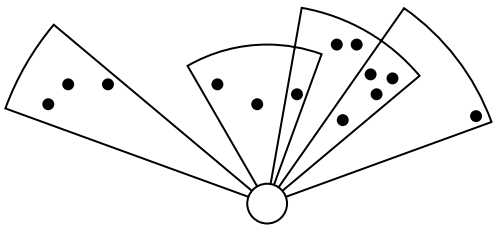}
\vspace{-4ex}
\end{center}
\caption{{\small The sectors correspond to the reaches
of directional antennas.}}
\label{fig1}
\end{figure}

We consider the following problem of optimal placement
of directional antennas in wireless networks.

There is a base station coupled with a network infrastructure.
The station transfers information to and from a number 
of customers within the range of directional antennas
placed at this station. Each customer has fixed position
and demand on the transmission capacity. The demands
are unsplittable, thus a customer can be assigned
only to a single antenna. One can choose the orientation
and the angular range of an antenna. When the angular range
is narrower an antenna can reach further so the area covered
by any antenna is always the same. There is a 
common limit on the total 
bandwidth demand that can be assigned to an antenna. 
The objective is to minimize the number of antennas.

Berman et al. termed this problem as
\MAV and provided
an approximation polynomial-time algorithm
with ratio $3$ \cite{B07}.
They also observed in \cite{B07} that even when
the angular range of antennas is fixed,
\MAV cannot be approximated 
in polynomial time with ratio
smaller than 1.5 by a straightforward
reduction from \Parti (see \cite{GJ}).

We provide a substantially better polynomial-time
 approximation algorithm
for \MAV achieving the ratio of $2.357.$ 
Our algorithm is based on two new
results which
are of independent interest in their own rights.

The first of these results states that a
cover of the set of customers with the minimum number of antennas
without the demand constraint can be
found in polynomial time. Previously,
only a polynomial-time approximation
with ratio $2$ as well
as an integrality gap with set cover ILP
were established for this
problem in \cite{B07}.

The second result shows that generally, given
an approximate solution
with ratio $r$ to an instance
of (uncapacitated) set cover, one can
find a solution to a corresponding
instance of the capacitated set cover,
where each set has the same capacity,
within $r+1.357$ of the optimum.

Berman et al. considered also
the following related problem which they
termed as \Bab \cite{B07}. There is a number
of items to be delivered. The $i$-th item
has a weight $d_i$, arrival time $t_i$
and patience $p_i$, which means that it
has to be shipped at latest by $t_i+p_i.$
Given a capacity of a single shipment,
the objective is minimize the number 
of shipments.

Similarly as Berman et al. could adopt their
approximation for \MAV to obtain
an approximation with ratio $3$ for \Bab \cite{B07},
we can adopt our approximation for \MAV
to obtain a polynomial-time approximation
algorithm with ratio $2.357$ for \Bab.

\begin{figure}[ht]
\begin{center}
\includegraphics[scale=0.75]{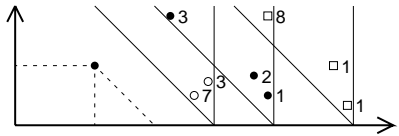}
\end{center}
\caption{\small{The $X$ coordinate of an item $i$ encodes $t_i$
and the $Y$ coordinate encodes $p_i$.
Shipment has capacity 10. The numbers 
indicate the weights. Items which are to
be shipped together must be enclosed by an angle.}}
\label{fig2}
\end{figure}

Our third main result 
is a PTAS for a dual problem to capacitated set cover
where
the number of sets (e.g., antennas) to use is fixed and the
task is to minimize the maximum set load, in case
the sets correspond to line intervals or arcs.
In the application to directional antennas,
the aforementioned correspondence comes from
fixing the radius and hence
also the angular range of the antennas and the 
problem has been termed as \MAb in \cite{B07}.
The task is to minimize the maximum load of an antenna.
In \cite{B07}, there has been solely presented
a polynomial-time approximation with ratio $1.5$ for \MAb.

\noindent{\textbf{Organization:}}   In Section \ref{firs} we present problem definitions and notations. In Section \ref{sec: uncap}, we derive our
polynomial-time dynamic programming method for the uncapacitated
variant of \MAV . In Section \ref{sec: cap}, we show our general
method of the approximate reduction of the capacitated
vertex cover to the corresponding uncapacitated one.
By combing it with the method of Section \ref{sec: uncap},
we obtain the $2.357$ approximation for \MAV.
Finally, in Section \ref{sec: dual}, we present the
PTAS for \MAb, or more generally, for
minimizing the maximum load in capacitated
set cover of bounded cardinality, in case
the sets correspond to intervals or arcs.

\section{Preliminaries} \label{firs}

This section presents terminology and notation used throughout this paper.

We use $U$ to denote $\{1,2,\ldots,n\}$.  If $x_i\in{\mathbb R}$ are defined
for $i\in U$ and $A\subset U$, $x(A)=\sum_{i\in A}x_i$.

An instance of the set cover problem is given by family \Sc of subsets of
$U=\{1,\ldots,n\}$; a cover is $\cC\subset\cS$ such that $\bigcup_{A\in\cC}A=U$.
We minimize $|\cC|$.  An instance of capacitated set cover also specifies $d_i$
for $i\in U$; a capacitated cover is a family of sets $\cC$ such that
(i) for each $A\in\cC$ there exists $B\in\cS$ s.t. $A\subset B$,
while $d(A)\le 1$; (ii) $\bigcup_{A\in\cC}A=U$.  Again, we minimize $|\cC|$.

If for each $j\in U$ we define radial coordinates $(r_j,\theta_j)$, we
define angle sector with radius bound as

\vspace{-3.5ex}
\[ \cR (r,\alpha,\delta)=\{ j \in U:~
r_j \leq r \mbox{ and }  \theta_j=\alpha+\beta  
\mbox{ with } 0\le\beta\le\delta \} \vspace{-1ex}  .\]

In \MAV as well as its uncapacitated variant, $U$ is the set of customers with
radial coordinates defined in respect to the position of the base station.
This is a variant of capacitated (or uncapacitated)
set cover where \Sc consists of sets of
customers that can be within range of a single antenna, \IE of the form
$\cR(r,\alpha,\rho(r))$, where $\rho(r)$ is the angular width of
an antenna with radial reach $r$.

The trade-off function $\rho$ is decreasing;
to simplify the proofs, we assume that $\rho(r)=1/r$, we can change the
$r$-coordinates to obtains exactly the same family of antenna sets as for
arbitrary $\rho$.

\section{Uncapacitated cover by antenna sets}\label{sec: uncap}

To simplify proofs,
we will ignore the fact that the radial coordinate has a ``wrap-around''.
We also renumber the customers so $\theta_i<\theta_{i+1}$ for $1\le i < n$.
Observe that if $\theta_i=\theta_j$ and $r_i\ge r_j$ then every
antenna set that contains $i$ also contains $j$, so we can remove $j$
from the input.

It suffices to consider only $n(n+1)/2$ different
antenna sets. For such an antenna set $A$, let $i=\min A$, $j=\max A$.
If $i=j$, 
we denote $A$ as $A[i,i]=\{i\}$, and
if $i<j$, we set $r(i,j)=(\theta_j-\theta_i)^{-1}$ and define
$A[i,j]=\cR(r(i,j),\theta_i,1/r(i,j))$.
(This definition is more complicated when the ``wrap-around'' is allowed.)
Because $A\subseteq A[i,j]$ we can use $A[i,j]$ in our set cover instead of $A$.

We say that points $i$ and $j$ are compatible, denoted $i\compat j$,
if $i\le j$ and there exists an antenna set that contains $\{i,j\}$.
If $i=j$ then $i\compat j$ is obvious; if $i<j$ then
$i\compat j\equiv \{i,j\}\subseteq A[i,j]\equiv r_i,r_j\le r(i,j)$.
If $i\compat j$, we define $S[i,j]=\{k:~i\le k\le j\} \setminus A[i,j]$.

We solve our minimum cover problem by dynamic programming.  Our recursive
subproblem is specified by a compatible pair $i,j$ and 
its objective is to compute the
size of minimum cover $C[i,j]$ of $S[i,j]$ with antenna sets.  
If we modify the input by
adding the points $0$ and $n+1$ with coordinates $(\theta_1-1,\varepsilon)$ and
$(\theta_n+1,\varepsilon)$ then our original problem 
reduces to computing $C[0,n+1]$.

If $S[i,j]=\varnothing$ then $C_{i,j}=0$.
Otherwise, $S[i,j]=\{a_0,\ldots,a_{m-1}\}$, where $a_k<a_{k+1}$
for $k=0,\ldots,m-2$.
  
We define a weighted graph $G_{i,j}=(V_{i,j},E_{i,j},c)$, where
$V_{i,j}=\{0,\ldots,m\}$, $(k,\ell+1)\in E_{i,j}$ iff
$a_k\Compat a_\ell$ and for an edge $(k,\ell+1)$ we define
the cost $c(k,\ell+1)=1+C[a_k,a_\ell]$.

Note that
$G_{i,j}$ is acyclic. Therefore,
we can find a shortest (i.e., of minimum total cost)
path from $0$ to $m$ in time
$O(|E_{i,j}|)=O(n^2)$ \cite{CLR}.
Let $d$ be the length of this path.  We will argue that $C[i,j]=d$.

First, we show a cover of $S[i,j]$ with $d$ antenna sets.
A path from 0 to $m$ in $G_{i,j}$ is an increasing sequence, and a path edge
$(u,v)$ with cost $c$ corresponds to a cover of $\{a_u,a_{u+1},\ldots,a_{v-1}\}$
with $A[a_{u},a_{v-1}]$ and $c-1$ antenna 
sets that cover $S[a_{u},a_{v-1}]$.

Conversely, given a cover \Cc of $S[i,j]$, we can obtain 
a path with cost $|\cC|$ in $G_{i,j}$ that connects $0$ with $m.$ 

For $A[k,\ell]\in\cC ,$ we say that $\ell-k$ is its {\em width}.
To make a conversion from a
cover \Cc of $S[i,j]$
to a path in $G_{i,j}$, we request that \Cc has
the minimum sum of widths among the minimum covers of $S[i,j]$.

This property of \Cc implies that if $A[k,\ell]\in\cC$ then:
\begin{enumerate}
\vspace{-1ex}
\item[] $k,\ell\in S[i,j]$,
\vspace{-1ex}
\item[] $k$ and $\ell$ are not covered by $\cC-\{A[k,\ell]\}$
(otherwise we eliminate
$A[k,\ell]$ from \Cc or replace it with a set that has a smaller width).
\end{enumerate}
\vspace{-1ex}

From this we can conclude that for each pair of sets
$A[k,\ell], A[k',\ell']\in\cC$, where $k<k'$,
 one of two following cases applies:
\begin{enumerate}
\vspace{-1ex}
\item
$\ell < k'$, i.e.,  $A[k,\ell]$ precedes $A[k',\ell']$;
\vspace{-1ex}
\item
\vspace{-1ex}
$\ell'<\ell$, i.e.,  $A[k',\ell']$ is nested in $A[k,\ell]$.
\end{enumerate}
\vspace{-1ex}

Let \Dc be the family of
those sets in \Cc that are not nested in others. Clearly
\Dc can be ordered by the leftmost elements in the sets.
Note that if $A[k,\ell]\in\cD$ then 
for some $f,g,c$, we have

\begin{enumerate}
\vspace{-1ex}
\item[] $a_f=k\in S[i,j]$,

\vspace{-1ex}
\item[] $a_g=\ell\in S[i,j]$,

\vspace{-1ex}
\item[] $c-1$ sets of \Cc are nested in $A[k,\ell]$ and they cover $S[i,j]$,

\vspace{-1ex}
\item[] $(f,g+1)$ is an edge in $G_{i,j}$ with cost $c$,

\vspace{-1ex}
\item[] $g+1=m$ or $A[a_{g+1},\ell']\in\cD$ for some $\ell'$.
\end{enumerate}
\vspace{-1ex}

These $(f,g+1)$ edges form a path that connects $0$ with $m$ with cost $|\cC|$.

Our dynamic programming algorithm solves the
$n(n+1)/2$ subproblems specified by compatible pairs
$i, j$ in a non-decreasing order of the differences
$j-i.$ In the reduction of a subproblem to
already solved subproblems the most expensive
is the construction of the graph $G_{i,j}$ and
finding the shortest path in it, both take quadratic
time. Hence, we obtain our main result in this section.

\begin{theorem}
The uncapacitated version of the problem of minimum
covering with antenna sets $n$ points, i.e., the restriction
of \MAV to the case where all point demands are zero,
can be solved in time $O(n^4)$ and space $O(n^2).$
\end{theorem}

\vspace{-1ex}
Previously, only a polynomial-time approximation algorithm
with ratio two was known for the uncapacitated version
of \MAV \cite{B07}.

\vspace{-1ex}
\section{From set cover to capacitated set cover}\label{sec: cap}

By the discussion in the previous section, it is sufficient to consider only
$O(n^2)$ antenna sets in an instance of \MAV on
$n$ points. Hence, \MAV is a special case of minimum capacitated set cover.
 
Since we can determine a minimum
uncapacitated set cover of an instance 
of \MAV by ignoring the demands
and running the dynamic programming
method given in the previous section,
we shall consider the following
more general situation.

We are given an instance of the general
problem of minimum capacitated set cover
and a minimum set cover of the
corresponding instance of minimum
set cover obtained by removing
the demands. The objective
is to find a good approximation
of a minimum capacitated set cover
of the input instance.

\vspace{-1ex}
\subsection{Approximation ratio $r+1.692$}\label{sec: 2.692}

We obtain an approximation with ratio 2.692
for minimum capacitated set cover on the
base of minimum uncapacitated
set cover \Ucop by running a simple greedy FFD
algorithm (see Fig. \ref{greedy}).
Our analysis of this algorithm in part resembles
that of the first-fit heuristic for bin-packing
\cite{Cof,GGJ76}, but the underlying problems are different.

\newcommand{\bec}{\leftarrow}
\newcommand{\ffor}{{\bf for}\ }
\newcommand{\iif}{{\bf if}\ }
\newcommand{\wwhile}{{\bf while}\ }
\begin{figure}
\begin{tabbing}
123456\= 1234\= 1234\= 1234\= 1234\= \kill
\>$\cQ\bec\es$\\
\>\ffor ($U\in\cUop$)\\
\>\> \wwhile ($U\not=\es$)\\
\>\>\> $Q \bec \es$\\
\>\>\> \ffor ($i\in U$, with $d_i$ non-decreasing)\\
\>\>\>\> \iif ($d(Q)+d_i\le 1)$\\
\>\>\>\>\> insert $i$ to $Q$ \\
\>\>\>\>\> remove $i$ from $U$ and $P$\\
\>\>\> insert $Q$ to \Qc \\[-4ex]
\end{tabbing}
\caption{\label{greedy}FFD, First Fit Decreasing algorithm for
  converting a cover into a capacitated cover.}
\end{figure}

\begin{theorem}\label{lem: 2.692} 
Given an instance of capacitated set cover 
on $n$ elements and an approximation
with ratio $r$ for minimum set cover
of the uncapacitated version of
the instance obtained by removing
the demands, a capacitated set cover of
the input  instance of size at most $r+1.692$ times larger than
the optimum can be determined in time $O(n^2).$ 
\end{theorem}

\begin{proof}
To analyze FFD,
we introduce a ``slack function'' $s(x)$, and we also apply it to
elements that we cover using notation notation $s_i=s(d_i)$.
Slack function has the following two properties:

\begin{dingautolist}{172}
\vspace{-1ex}
\item
if $d(Q)\le 1$ then $s(Q)\le 0.692$;
\vspace{-1ex}
\item
if \wwhile($U\not=\es$) loop produces $\ell+1$ solution sets,
say $Q_0,\ldots,Q_\ell$ then $\sum_{j=0}^\ell(d(Q_j)+s(Q_j))\ge \ell$.
\end{dingautolist}
\vspace{-1ex}

Let \Qcop be the optimum solution.
Property \ding{172} implies that we start with $s(P)\le 0.692|\cQop|$.
Property \ding{173} implies that algorithm FFD produces at most
$|\cUop|+d(P)+s(P)\le 2.692|\cQop|$ sets.
It remains to prove \ding{172} and \ding{173}.

We define intervals 
$\cI_k=\{x:\frac{1}{k+1}<x\le\frac{1}{k}\}$, $k=1,2,\ldots$,
and we use them to divide
$P$ into classes,
$P_k=\{i\in P:~d_i\in\cI_k\}$.
Now we define the slack function:
$ s(x)=\frac{1}{k(k+1)} \mbox{if } x\in\cI_k$.

We also introduce $r(x)=s(x)/x$ and $r_i=r(d_i)$; observe that

$s(Q)\le d(Q)\max_{i\in Q}r_i$;

$\frac{1}{k+1}\le r_i <\frac{1}{k}$ for $i\in P_k$.

To prove \ding{172}, we look for the maximum possible $s(Q)$.
If $d(Q)\le 1$ and $s(Q)>\fotw$, then for some ${a_0}\in Q$ we have
$r_{a_0}\ge\fotw$, hence ${a_0}\in P_1$, so
$d_{a_0}=\fotw+\eps$ for some $\eps>0$ and $s_{a_0}=\fotw$.

It remains to find maximum possible $s(Q^1)$ where $Q^1=Q-\{a_0\}$.  Note that
$d(Q^1)=\fotw-\eps$, (thus $Q\cap P_1=\es$).
If $s(Q^1)\ge\frac{1}{6}$ then for some 
$a_1\in Q^1$ we have $r_{a_1}\ge\fotr$, hence ${a_1}\in P_2$ and
$d_{a_1}=\fotr+\eps$ for some $\eps>0$.

We can repeat the reasoning with $Q^2=Q_1-\{a_1\}$ and conclude
that it contains $a_2\in P_6$, and then with $Q^3=Q^2-\{a_2\}$ we
can conclude that it contains $a_3\in P_{42}$, etc.  Subsequent
terms contribute very little to the overall result, so we can approximate
the maximum possible $s(Q)$ as
$\fotw+\frac{1}{6}+\frac{1}{42}+\frac{1}{42\times 43}\approx 0.69103$.

The proof of property \ding{173} is in Appendix A.

Since $|$ \Ucop $|\le n,$ our simple algorithm can be
implemented in time $O(n^2).$ 
\end{proof}

\vspace{-5ex}
\subsection{Approximation ratio $r+1.423$}                     

FFD algorithm achieves the worst case behavior if the sets of
the optimum solutions have demands of the form
$\{\frac{1}{2}+\eps,\frac{1}{3}+\eps,\frac{1}{7}+\eps,\frac{1}{43}+\eps,\ldots\}$
and the uncapacitated cover \Ucop has sets that either have very small
$d(U)$, or group together all elements with a particular weight.

E.g., for $U$ that contains elements with $d_a=\fotw+\eps$, algorithm FFD
creates one-element sets.  We can improve the approximation by preceding
FFD with a phase in which we attempt to create ``better sets''.

If $d(Q)\le 1$ and $Q\cap P_1=\es$, the maximum $s(Q)$ is obtained
with demands
$\frac{1}{3}+\eps$,
$\frac{1}{3}+\eps$,
$\frac{1}{4}+\eps$,
$\frac{1}{13}+\eps,\ldots$,
and this yields $s(Q)=
\frac{1}{6}+
\frac{1}{6}+
\frac{1}{12}+
\frac{1}{156}+\ldots\approx 0.4231$.

We can achieve the same even if there exists $a\in Q\cap P_1$ if
we reduce $s_a$ from $\fotw$ by about $0.269$, to about $0.231$.  
Then we need to modify the algorithm so it produces sets with $d(Q)+s(Q)\ge 1$.
This is not necessarily possible, after all, \Qcop may even contain
singleton sets.  For this reason, we add the third term to our
amortization of sets.  For $a\in P_1$ we define

$Q_a$ is the set in \Qcop such that $a\in Q_a$;

$x_a=1-d(Q_a)$;

$y_a=0.1905-s(Q_a-\{a\})$.

For $a\not\in P_1$ we set $x_a=0$.  Clearly, $x(P)+d(P)\le|\cQop|$
while $s(P)+y(P)\le 0.4222|\cQop|$.  Thus it will suffice to produce
sets such that $(d+x+s+y)(Q)\ge 1$, and for that, we just need to
modify the way we create sets that contain elements of $P_1$.

Let us consider what we (nondeterministically) can do, and what we need to do.
Consider $a\in P_1$ and assume that $d_a-\fotw=x_a=y_a=0$.  Then
we can find $S\in\cS$ and $A\subset S$ such that $a\in S$, $A\subset P-P_1$
and $s(A)=0.6905$.  However, it suffices to find $A$ such that
$s(A)=0.269$, less than 40\% of what we can do.

If we increase $d_a$, $x_a$ or $y_a$ by some $\delta$, both what we can do
and what we should do decrease by $\delta$, hence the ratio decreases.

We can find a good candidate for $A$ by ``guessing''
$S\in\cS$ and
running an approximation algorithm for the knapsack problem \cite{KP04}
in which items are elements $i\in S-\{a\}$, the weights are $d_i$,
the values are $v_i=d_i+s_i$.  It suffices to have 80\% approximation.

When we find a set $A_a$ that has the maximum value (as returned by the
approximation algorithm), we form set $B_a=A_a\cup\{a\}$.  We do the following
``accounting trick''.  For each $i\in A$ and $b\in P_1$, if $b\not=a$
and $i\in Q_b$, then we increase $x_b$ by $\fotw v_i$.
Thus we achieve $d_a+s_a+x_a+y_a+\fotw v(A)\ge 1$, while for the
remaining elements $b\in P_1$ the ratio of what ``they can do'' (maximum
possible $v(A_b)$) to what ``they need to do'' (the difference
$1-d_b-x_b-y_b$) remains bounded by 40\%.

After creating $B_a$ for each $a\in P_1$ we run FFD algorithm with
the remaining elements.

In this preliminary version we omit details how to implement this
refined algorithm in time $O(n|\cS|)$.


\begin{theorem}\label{lem: 2.423} 
Given an instance of capacitated set cover 
on $n$ elements and an
approximation with ratio $r$
for minimum set cover
of the uncapacitated version of
the instance obtained by removing
the demands, a capacitated set cover of
the input  instance of size at most $r+1.423$ times larger than
the optimum can be determined in polynomial time. 
\end{theorem} 

\subsection{Approximation ratio $r+1.357$}                     

One can observe that algorithm FFD has worst performance when
some peculiar combinations of demands occur in sets of the optimum
solutions, in terms of our classes, the worst pattern
is $(P_1,P_2,P_6,\ldots)$.  Our second algorithm has an initial phase
that handles all sets with an element from $P_1$; we decrease the
slack for elements of $P_1$ and spend more effort forming the
sets, so even with the smaller slack we can amortize the cost
of each set of our solution.

Intuitively, members of $P_1$ were troublemakers and our added phase
took care of that.

Because knapsack problem has fully polynomial-time approximation schema
we could run a version with, say, 99\% approximation, and this would
allow to decrease the slack in $P_1$ by almost $0.6903/2$.
This would give an approximation ratio of about $2+0,7/2=2.35$.
However, at this point we get another worst case --- with the pattern
$(P_2,P_2,P_3,P_1,\ldots)$.

We say that $a\in P_2$ is a troublemaker if for some $Q$ we have
$a\in Q\in\cQop$ and $|Q\cap P_2|=2$.  Here both elements
of $Q\cap P_2$ are troublemakers, we call them siblings.

Now we will describe how to add a second phase to the algorithm
so that the case of sibling troublemakers will cease to be the worst
one.  At that point we will have two classes of worse cases: $(P_1,\ldots)$,
because they are compatible only with approximation ratios that
are at least $2.35$, and $(P_2,P_3,\ldots)$.  The worst of the
latter is $(P_2,P_3,P_3,P_7,\ldots)$.  One can see that the slack
of the latter is almost like the slack of the worst case of FFD, except
that we have replaced a demand from $P_1$ with two from $P_3$, $\fotw+\eps$
with two $\frac{1}{4}+\eps$.  Thus this slack is approximately
$0.6903-0.3333=0.357$.

The second phase is similar to the first phase:
we ``guess'' a set $S\in\cS$, elements $a,b\in S\cap P_2$ and we
run an approximation algorithm to find $B\in S-\{a,b\}$ such that
$d(B)\le 1-d_a-d_b$, while we maximize $s(B)$.  For all possible
guesses, we pick one with maximum $d(B)+d_a+d_b$, form set 
the $Q=B\cup\{a,b\}$,
insert $Q$ to our solution and remove $Q$ from $P$.
We repeat it as long as there exists $S\in\cS$ with $|S\cap P_2|\ge 2$.

After the second phase is completed, we finish by running FFD with
the remaining $P$, the set of still uncovered elements.

To analyze the second phase we introduce a negative slack for each pair
of sibling troublemakers, 0.1.  When we form a set that contains
troublemakers, we amortize it with the sum of the demands and slacks
of elements, plus the slacks (and extra terms) of the troublemaker sibling
pairs that are involved.

One can see that the sum of slacks in $Q\in\cQop$ that has a pair $a,b$
of troublemakers is at most $0.323$ --- we specifically decreased it
by $0.1$.  We also define the extra terms similarly as before:

$x_{a,b}=\fotr-d(Q-\{a,b\})$;

$y_{a,b}=0.423-s(Q)$.

If $x_{a,b}+y_{a,b}=0$, then the pair $a,b$ ``needs to find'' 0.1, and
it ``can find'' 0.423, so it suffices if it finds 25\% of what it can
find.  When $x_{a,b}$ (or $y_{a,b}$) is positive, it decreases the
''need to find'' and ''can find'' by the same amount, so the ratio
only improves (decreases).

Now suppose that we form a set, and in the competition of ``guesses''
the winners were some $a,b\in P_2$.  The critical case is when they
are both troublemakers, each with its sibling, $a'$ and $b'$ respectively,
and needs, $N_a$ and $N_b$.  Because $a,a'$ could find $4N_a$,
$b,b'$ could find $4N_b$, they could find at least the average,
$2(N_a+N_b)$.  By applying
$\frac{2}{3}$ approximation, they found at least $\frac{4}{3}(N_a+N+b)$,
the use $\frac{3}{4}$ of that to satisfy their needs, and
$\frac{1}{4}$ of that to compensate the troublemakers whose
now can find less.  The compensated troublemakers maintain
their 25\% ratio of need/can.


In this way, we obtain our strongest approximation results.

\begin{thm} 
Let an instance of capacitated set cover be specified
by a universe set $P=\{1,...,n\}$, 
demands $d_i\ge 0$ for each $i\in P$,
and a family \Sc of subsets of $P.$
If an approximation
with ratio $r$ for minimum set cover of the uncapacitated
version of the instance (i.e., where the demands
are removed) is given then a capacitated set cover of
the input  instance of size at most $r+1.357$ times larger than
the optimum can be determined in polynomial time. 
\end{thm}

\begin{cor}
There exists a polynomial-time approximation algorithm for the problem
of \MAV with ratio $2.357$.
\end{cor}

By the reduction of \Bab to \MAV given in \cite{B07},
we also obtain the following corollary.

\begin{cor}
There exists a polynomial-time approximation algorithm for the problem of \Bab with ratio $2.357$.
\end{cor}

\section{PTAS for \MAb}
\label{sec: dual}

In \MAb problem, the radius of antennas is fixed and
the number $m$ of antennas that may be used is specified.
The task is to minimize the maximum load of an antenna.
In \cite{B07}, there is presented
a polynomial-time approximation with ratio $1.5.$

In the dual problem \MAa, the maximum load is fixed
and the task is to minimize the
number of antennas.  
Recall that achieving
an approximation ratio better than $1.5$
for the latter problem  requires solving the 
following problem equivalent to 
{\sc Partition}.

Suppose that all demands can be covered with a single set, the
load threshold is $D$ and 
the sum of all demands is to $2D$.
Decide whether or not two antennas are sufficient
(which holds if and only 
if one can split the demands into two equal parts).  

However, in case of the corresponding instance of \MAb ,
we can apply FPTAS for the {\sc SubsetSum} problem \cite{KPS03}
in order to obtain
a good approximation for the minimization of the larger of the two loads.

If all demands can be covered by a single antenna set (and the sum of demands
is arbitrary) then \MAb problem is equivalent
to that of minimizing the makespan
while scheduling jobs on $m$ identical machines.  Hochbaum and Shmoys showed
a PTAS for this case in \cite{HS87}.

Interestingly enough, the PTAS of Hochbaum and Shmoys can be modified for \MAb,
while it does not seem to be the case with their practical algorithms that have
approximation ratios of 6/5 and 7/6 \cite{HS87}.

Because radial coordinate does not matter in \MAb, the input is a sequence of
pairs $(\theta_i,d_i),~i=1,\ldots,n$.  Initially, we ignore the issue of
``wrap-around'' so the antenna sets are of the form
$\mathcal{R} (\alpha)=\{j\in U:~\alpha\le\theta_j<\alpha+\Theta\}$.
Without loss of generality we assume that $U=\{1,\ldots,n\}$ and
$\theta_1<\theta_2<\ldots<\theta_n$.

In our PTAS, we try different values of the maximum load $D$. We can
start using simple factor 2 approximation and then we can apply binary search.
We will find an exact solution for a transformed problem in such a way that
(a) the cost of the optimum cannot increase, (b) a solution for the
transformed problem can be converted to an actual solution while increasing
the cost by a factor of $1+\eps$.

For a fixed $k,$ we will describe an 
$(1+\epsilon)$-approximation algorithm that runs in time $O(n^{k+c})$,
where $c$ is a universal constant, while $\eps\approx(1+\ln k)/k$.

We start by defining thresholds $t_i=D(1+\eps_0)^{-i}$ and classes:\\
$C_i=\{j\in U:~t_{i+1}\le d_i<t_i\}$, $i=0,\ldots,k-1$
({\em large} demands) and\\
$C_k=\{j\in U:~d_i<t_k\}$ ({\em small} demands).
We also set $\eps_1=t_k$ and $\eps=\eps_0+\eps_1$.  One can show 
that $\eps$ is minimized when $\eps_1\approx 1/k$ and $\eps_0\approx \ln k/k$.

\newcommand{\decrel}{{\sc Decreased}\xspace}
We will find exact solution to a problem where we
have the same input but we re-define the cost/load
of sets so (a) it cannot decrease and (b) if the new
cost of $Q$ satisfies $cost(Q)\le D$ then $d(Q)\le(1+\eps)D$.
We call this problem {\sc Decreased}.

Intuitively, we divide elements into small and large.  In the
case of large elements, with $d_j>t$, we decrease $d_j$ to
$d'_j$ to have a small number of distinct values.  In the case
of small elements, we want to apply ``greedy packing'' and we
``decrease'' their contribution by not counting the last of them.
More formally,
we define decreased/relaxed instance \decrel as follows:

for $j\in C_i$, we set $d'_j$ to $t_{i+1},$ 

if $Q\cap C_k=\es$, we set $cost(Q)$ to
$d'(Q),$ i.e., $\sum_{j\in Q}d'_j$, otherwise

if $j=\max(Q\cap C_k)$, we set $cost(Q)$
to $d'(Q-C_k)+d(Q\cap C_k)-d_j,$

the task is to minimize $\max_{Q\in\cQ}cost(Q)$.

Clearly, the optimum of our \decrel instance cannot be larger than
the optimum for the initial \MAb instance.
Also, since if $j\in C_i$ for $i<k$ then $d_j\le (1+\epsilon_0)d'_j$
and otherwise $d_j\le d'_j +\epsilon$
we conclude that
$d'(Q)\le D$ implies $d(Q)\le (1+\epsilon)D.$
Thus, an exact polynomial-time algorithm for \decrel
yields a PTAS for \MAb.

\noindent
We say that a partition $\cQ$ of $U$ is {\em ordered} if we have the
following implication: if $Q,Q'\in\cQ$, $\max(Q)<\max(Q')$,
$j\in Q\cap C_i$, $j'\in Q'\cap C_i$, then $j<j'$.
\begin{lemma}
For every solution $\cQ'$ of \MAb there exists
an ordered solution $\cQ$ of \decrel such that
$\max_{Q\in\cQ}cost(Q)\le\max_{Q\in\cQ'}d(Q)$.
\end{lemma}
\begin{proof}
We can transform $\cQ'$ to an ordered $\cQ$ in such a way that during that
process for every $Q\in\cQ'$ we will preserve $|Q\cap C_i|$ for each
$i>k$ and we will not increase $d(Q\cap C_k)$.  Before $Q$ is ``finalized''
we will allow fractional values for statements $[j\in Q]$ if $j\in C_k$.

Consider $Q\in\cQ'$ that has minimal $max(Q)$ and suppose that there exists
$Q'\in \cQ'-\{Q\}$ and $j,j'\in C_i$, $j<j'$ such that $[j\in Q']>0$ and
$[j'\in Q]>0$.  If $i<k$, we move $j$ to $Q$ and $j'$ to $Q'$;
this does not change $cost(Q-C_k)$ and $cost(Q'-C_k)$.
If $i=k$, let
$x=\min \{[j\in Q'],[j'\in Q]\}$, 
we increase $[j\in Q]$ and $[j'\in Q']$ by $x$
and we decrease $[j'\in Q]$ and $[j\in Q']$ by the same amount.
This does not change $d(Q\cap C_k)$ and $d(Q'\cap C_k)$.

When such $Q',i,j,j'$ do not exists, suppose that there exists
$j\in C_k$ such that $0<[j\in Q]<1$; in this case $j=\max(Q\cap C_k)$;
we increase $[j\in Q]$ to 1 and for $Q'\not=Q$ we decrease
$[j\in Q']$ to 0.  This does not increase $cost(Q)$ because $cost$ does not
count the last small element in $Q$.

Now $Q$ and any other $Q'$ satisfy the condition of {\em ordered} and we can
remove $Q$ and its elements from further consideration---and insert
$Q$ to $\cQ$.
We repeat this until all sets are removed from $\cQ'.$
\end{proof}

The algorithm based on the lemma can be as follows.  We represent a partial
solution as counts $(c_0,\ldots,c_k)$, that mean $c_i$ elements of class $C_i$
were covered.
The are at most
$\Pi_{i=0}^k|C_i|\le
(n/(k+1))^{k+1}$ such partial solutions.
Because we add sets to a solution in order of increasing $\max(Q)$,
a partial solution covers $c_i$ smallest elements of $C_i$ --- smallest
in terms of their $j$'s, or, equivalently, $\theta_j$'s.

Adding a set to a partial solution $(c_0,\ldots,c_k)$
is an edge to another such vector, $(c'_0,\ldots,c'_k)$.
Such an edge is determined by the sequence $(c'_0,\ldots,c'_{k-1})$,
because then we can find maximum possible $c'_k$.  An edge is valid
if it implies the increase in the maximum index of a covered element,
and $\sum_{i=0}^{k-1}(c'_i-c_i)t_{i+1}\le D$.  Because a new set can
cover at most $1/t_k=k$ large demands, the number of possible edges
is below $4^k$.  We need to find the shortest path from
$(0,\ldots,0)$ to $(|C_0|,\ldots,|C_k|)$, and we can use breadth first
search; thus the time is proportional to the number of edges, or
$O((4n/k)^{k+1})$. By $\eps\approx(1+\ln k)/k$, the time
can be also expressed as $n^{\frac 1{\eps}\ln \frac 1{\eps}+O(1)}.$
Hence, we obtain our PTAS for \MAb.

\begin{theorem}\label{theo: PTAS}
\MAb for $n$ points
admits an approximation with ratio $1+\eps$
in time $n^{\frac 1{\eps}\ln \frac 1{\eps}+O(1)}.$
\end{theorem}

Note that the only geometric property of antennas
with fixed radius that we used to design
the PTAS for \MAb is their correspondence
to intervals or arcs. Hence, we obtain
the following generalization of
Theorem \ref{theo: PTAS}. 

\begin{theorem}\label{theo: gPTAS}
The problem of minimizing the maximum
load in a capacitated set cover where
the sets correspond to intervals
or arcs admits a PTAS.
\end{theorem}


\section{Concluding Remarks}


We are quite convinced
that our general method of approximating with ratio $r+1.357$
minimum capacitated
set cover on the base of an approximate solution with ratio $r$
to the corresponding
minimum (uncapacitated) set cover can ultimately
achieve the ratio $r+1.3.$
In particular, this would improve the ratio for
\MAV to $2.3.$ It seems however that 
some new ideas are needed to obtain, if possible,
ratios below $r+1.3$ and $2.3,$ respectively.

Our aforementioned
method can be also used to approximate optimal
solutions to the natural extension of \MAV
to include several base stations by combining
it with known approximation algorithms
for geometric set cover in the plane, 
e.g., \cite{goodrich,clarkson,hoch1}.


\section{Acknowledgments}

The authors are grateful to 
Matin Wahlen for discussions on
\MAb and to David Ilcinkas,
Jurek Czyzowicz and Leszek Gasieniec
for preliminary discussions on \MAV.


\newpage
\section*{APPENDIX A: proof of the property \ding{173}}
\begin{dingautolist}{173}
\item
if \wwhile($U\not=\es$) loop produces $\ell+1$ solution sets,
say $Q_0,\ldots,Q_\ell$ then $\sum_{j=0}^\ell(d(Q_j)+s(Q_j))\ge \ell$.
\end{dingautolist}

We prove \ding{173} as follows.
We remove from consideration every set $Q$ created during that
loop if $d(Q)+s(Q)\ge 1$. For $j<\ell$ we can define positive deficit
$\delta_j=1-d(Q_j)-s(Q_j)$.  

The claim is trivial if $\ell=0$, \IE the loop creates only one set.  Moreover,
$d(Q_{\ell-1})+d(Q_\ell)>1$, hence it suffices
to show $s(Q_{\ell-1})+s(Q_\ell)+\sum_{j=0}^{\ell-2}(d(Q_j)+s(Q_j))>\ell-1$,
equivalently, $s(Q_{\ell-1})+s(Q_\ell)>\sum_{j=0}^{\ell-2}\delta_j$.

Let $t_j$ be the time when algorithm FFD initializes $Q_j\bec\es$;
and let $P_{L(j)}$ be the class of the largest element of $U$ time $t_j$.

If $|U\cap P_{L(j)}|\ge L(j)$ at time $t_j$, the algorithm would insert $L(j)$
elements of $P_{L(j)}$ to $Q_j$, as each $a\in P_{L(j)}$ satisfies
$d_a+s_a>\frac{1}{L(j)+1}+\frac{1}{L(j)(L(j)+1)}=\frac{1}{L(j)}$, this would
lead to in $d(Q_j)+s(Q_j)>1$; a contradiction because we removed such sets from
consideration.  Hence $|U\cap P_{L(j)}|<L(j)$ at time $t_j$ and the algorithm
inserts entire remaining $P_{L(j)}$ to $Q_j$ as well as at least one smaller
element.  This shows that $L(j)$ is increasing with $j$.

We will estimate the size of deficits and the ``surplus''
$s(Q_{\ell-1})+s(Q_\ell)$.

First, we estimate $s(Q_j)$ in terms of $\lambda=L(j+1)$.
While we form set $Q_j$, we can always insert an element from $P_\lambda$,
unless $1-d(Q)<\frac{1}{\lambda}$, so $Q_j$ has a subset $Q'$ with
$d(Q')>1-\frac{1}{\lambda}=\frac{\lambda-1}{\lambda}$ and
$\min_{a\in Q'}r_a\ge\frac{1}{\lambda+1}$, hence
$s(Q_j)>\frac{\lambda-1}{\lambda(\lambda+1)}=Est(\lambda)$.
$Est(\lambda)$ is decreasing with $\lambda$, starting with $\lambda=3$.
The case of $\lambda\le 2$ is not possible, because it implies that
$Q_j$ has an element of $P_1$, hence, $\delta_j<0$.

Second, we apply the same reasoning for
$Q_{\ell-1}\cup Q_\ell$
and $\Lambda=P(\ell)$: at time $t_\ell$ there exists
$b\in P_\Lambda$ and
$Q_{\ell-1}\cup\{b\}$ contains a subset $Q'$ such that $d(Q')>1$ and
$\min_{a\in Q'}r_a\ge\frac{1}{\Lambda+1}$, hence
$s(Q_{\ell-1})+s(Q_\ell)\ge s(Q')>\frac{1}{\Lambda+1}$.

Third, because we could insert $b$ when we were creating $Q_j$ for $j<\ell$
we have $d(Q_j)>1-\frac{1}{\Lambda}$.

Fourth, for $k>1$ we estimate $\delta_{\ell-k}$;
because $\lambda=L(\ell-k+1)\le\Lambda-k+1$, we have
$s(Q_{\ell-k})\ge Est(\Lambda-k+1)=\frac{\Lambda-k}{(\Lambda-k+1)(\Lambda-k+2)}$, hence 

$\delta_{\ell-k}\le 1-(1-\frac{1}{\Lambda})-
\frac{\Lambda-k}{(\Lambda-k+1)(\Lambda-k+2)}
=\frac{(\Lambda-k)(3-k)+2}{\Lambda(\Lambda-k+1)(\Lambda-k+2)}=est(k)$.

\vspace{0.5ex}\noindent
Because $\lambda\ge 3$, $\Lambda-k\ge 2$, this shows that we have positive
deficits only for
$k=2,3$ (for $k=1$ the estimate refers to $Q_{\ell-1}$ and this set contributes
to the surplus).  Thus it suffices to show that
$\frac{1}{\Lambda+1}-est(2)-est(3)\ge 0$:

\newcommand{\tmi}{-}
\newcommand{\tpl}{+}
$
\frac{1}{\Lambda\tpl 1}
-\frac{\Lambda\tmi 2\tpl 2}{\Lambda(\Lambda\tmi 1)\Lambda}
-\frac{2}{\Lambda(\Lambda\tmi 2)(\Lambda\tmi 1)}
=
\frac{1}{\Lambda\tpl 1}
-\frac{1}{(\Lambda\tmi 1)\Lambda}
-\frac{2}{(\Lambda\tmi 2)(\Lambda\tmi 1)\Lambda}=
$

$
\frac{1}{\Lambda\tpl 1}
-\frac{1}{(\Lambda\tmi 1)(\Lambda\tmi 2)}
$

\vspace{0.5ex}\noindent
In our fourth point of the reasoning we observed that $\Lambda-k\ge 2$,
and the smallest value of $k$ is 2, so $\Lambda\ge 4$ and the above estimate
is indeed positive.

\end{document}